\documentclass{elsart}
\usepackage{amssymb}
\usepackage[dvips]{epsfig}

\begin{document}

\begin{frontmatter}

\title{Random walk approach to the analytic solution of random systems with
multiplicative noise - the Anderson localization problem}

\author{V.N. Kuzovkov}

\address{Institute of Solid State Physics, University of
Latvia, \\ 8 Kengaraga Street, LV -- 1063 RIGA, Latvia}
\ead{kuzovkov@latnet.lv}

\author{ W.~von Niessen}

\address{Institut f\"ur Physikalische und Theoretische Chemie,
Technische Universit\"at Braunschweig, Hans-Sommer-Stra{\ss}e 10,
38106 Braunschweig, Germany}


\begin{abstract}
We discuss here in detail a new analytical random walk approach
to calculating the phase-diagram for spatially extended systems
with multiplicative noise. We use the Anderson localization
problem as an example. The transition from delocalized to
localized states is treated as a generalized diffusion with a
noise-induced first-order phase transition.  The generalized
diffusion manifests itself in the divergence of averages of
wavefunctions (correlators). This divergence is controlled by the
Lyapunov exponent $\gamma$, which is the inverse of the
localization length, $\xi=1/\gamma$. The appearance of the
generalized diffusion arises due to the instability of a
fundamental mode corresponding to correlators. The generalized
diffusion can be described in terms of signal theory, which
operates with the concepts of input and output signals and the
filter function. Delocalized states correspond to bounded output
signals, and localized states to unbounded output signals,
respectively. The transition from bounded to unbounded signals is
defined uniquely by the filter function $H(z)$.
\end{abstract}

\begin{keyword}
Random systems \sep diffusion \sep Anderson localization \sep
phase diagram

\PACS 05.40.-a \sep       64.60.-i \sep    72.15.Rn \sep 71.30.+h
\end{keyword}

\end{frontmatter}



\section{Introduction}

A variety of phenomena in physics, chemistry and biology is
modelled by stochastic differential (or difference) equations
\cite{Gardiner,VanKampen} which involve both additive and
multiplicative noise.
The additive noise is
relatively easy to handle with the help of the central limit
theorem. This type of noise disappears from the noise-averaged form
of the equations. The situation changes dramatically with the
appearance of multiplicative noise \cite{Qiang}. Some situations
have been documented in the literature, in which the multiplicative
noise participates in the creation of a pure \textit{noise-induced
phase transition} \cite{VandenBroeck}. In such a case fundamental
theorems of an importance comparable to the central limit theorem
are still lacking. In this paper we will investigate another famous
example of this phenomenon - the Anderson localization as a
noise-induced metal-insulator phase transition.

Let us consider spatially extended systems with multiplicative
noise. Anderson localization \cite{Anderson58} has remained a hot
topic in the physics of disordered systems for a long time (see
review articles \cite{Kramer,Janssen,Abrahams2}) and constitutes a
multidisciplinary problem.
Formally the Schr\"odinger equation with random on-site
potentials in the tight-binding representation is a stochastic
algebraic equation with multiplicative noise, where physical meaning
can only be attributed to certain average values (averaging over the
ensemble of random potentials \cite{Lifshitz}).
The problem is very similar to statistical physics, especially to
the statistical physics of phase transitions (quantum phase
transitions, see \cite{Vojta}), because there also a noise-induced
metal-insulator phase transition is analyzed (keywords: critical
dimension, correlation length, etc.). In the case of phase
transitions one finds that the relevant parameters (in our case
e.g. Lyapunov exponents or localization lengths) are not
analytical functions of the disorder.
Typically the theoretical analysis of disordered systems
is based either on approximations or on numerical methods
\cite{Kramer,Janssen}. However, it is clear that this cannot be
sufficient for systems revealing a \textit{phase transition}
\cite{Baxter}, e.g. a metal-insulator transition as is the case
with Anderson localization. In such a case an exact solution is
greatly needed.

Disordered systems are usually considered as a part of solid state
physics (or statistical physics). The theoretical methods here are
very different from the ones in the theory of dynamical systems.
It is known that static randomness (multiplicative noise in space)
in the physics of disordered systems may drastically change
macroscopic quantities. Such an effect exists also for dynamical
systems with randomness in the equations of motion (multiplicative
noise in time). For example, in special cases the random walk
problem is mapped to a quantum mechanical tight-binding model
exhibiting Anderson localization
\cite{Kuzovkov02,Kuzovkov04,Radons04}. It might be reasonable to
start from a simple microscopic stochastic kinematic model
(Brownian motion and normal diffusion), and to build the theory of
Anderson localization along this line (tight-binding model and
generalized diffusion \cite{Kuzovkov04}, with a noise-induced
metal-insulator phase transition).



In our papers \cite{Kuzovkov02,Kuzovkov04} we presented somewhat
schematically and by leaving out many mathematical details an
exact analytic solution of the Anderson localization problem.
Subsequently there were critical comments \cite{Comment} (see also
\cite{Reply}) that our proof is too short and condensed, and more
details are needed, the more so, since the mathematical formalism
used by us is quite new for this scientific community of Anderson
localization. This is why we present in the present paper the
detailed derivation of the analytical solution with illustrations.
Note that under the exact solution we mean the calculation of the
phase diagram for the metal-insulator system
\cite{Kuzovkov02,Kuzovkov04}. We do not calculate transport and
other important properties. However, the knowledge of the phase
diagram permits to understand, how these properties can be
calculated.


The structure of the present paper is as follows. In Section 2 we
explain how the localized states can be treated in terms of a
\textit{generalized diffusion}. This approach allows to understand
why for defining the phase diagram it is sufficient to solve
exactly only equations for the \textit{joint correlators}. We
present and solve here the equations for arbitrary dimensions.
Section 3 deals with the \textit{stability} of this solution. It
is shown that the generalized diffusion arises due to the
divergence of the \textit{fundamental mode}. The determination of
the stability range of this mode permits us to calculate the
\textit{phase diagram}. We explain our language of \textit{input
and output signals, filter function} which is new for the Anderson
localization community.

\section{Equations for correlators}

\subsection{General aspects}

Strictly speaking, the disordered Anderson tight-binding model
\cite{Anderson58} with Schr\"o\-din\-ger equation
\begin{equation} \label{tight-binding}
\sum_{\mathcal{M}^{\prime}}\psi_{\mathcal{M}^{\prime}}=(E-\varepsilon_{\mathcal{M}})\psi_{\mathcal{M}}
,
\end{equation}
where the summation over $\mathcal{M}^{\prime}$ runs over the
nearest neighbours of site $\mathcal{M}=\{m_1,m_2,\dots,m_D\}$,
cannot be solved exactly for arbitrary dimension D. The reason is
that the random potential $\varepsilon_{\mathcal{M}}$ enters the
equation as a product with the random amplitude
$\psi_{\mathcal{M}}$ which corresponds to the multiplicative noise
case \cite{Kuzovkov02}. Therefore, the \textit{exact solution} of
eq. (\ref{tight-binding}) discussed in
\cite{Kuzovkov02,Kuzovkov04} is possible only under special
circumstances to be discussed below.

The fundamental quantity of a disordered system - the localization
length $\xi$ - was determined in \cite{Kuzovkov02,Kuzovkov04} via
the Lyapunov exponent $\gamma$. In these calculations the following
conditions which appear contradictory at first sight have to be
satisfied. The phase diagram of the system with metal-insulator
transition should be obtained in the \textit{thermodynamical limit}
\cite{Baxter} (the infinite system). However, the determination of
the Lyapunov exponent needs the introduction of a coordinate system
(starting point) which imposes certain limitations on the system
size. Moreover, a direction for the growth of the divergent quantity
should be chosen, despite the fact that all space directions are
equivalent in eq. (\ref{tight-binding}).

All these conditions are fulfilled for the \textit{semi-infinite}
system, or system with a boundary, where the index $n \equiv
m_D\geq 0$, but all $m_j\in(-\infty,\infty)$, $j=1,2,\dots,p$,
with $p=D-1$. The boundary which is the layer $n=0$ defines the
preferential direction (the axis $n$) along which the Lyapunov
exponent $\gamma$ will be calculated.

It is convenient to interpret the index $n=0,1,\dots,\infty$ as
the \textit{discrete-time}, whereas all other indices combine in
the vector $\mathbf{m}=\{m_1,m_2,\dots,m_p\}$. The Schr\"odinger
equation (\ref{tight-binding}) can be rewritten as a
\textit{recursion equation} (in terms of the discrete-time, and
assuming summation over repeated indices)
\begin{equation} \label{recursion DN}
\psi _{n+1,\mathbf{m}}=-\varepsilon _{n,\mathbf{m}} \psi
_{n,\mathbf{m}}-\psi_{n-1,\mathbf{m}} +
\mathcal{L}_{\mathbf{m},\mathbf{m^{\prime}}}\psi
_{n,\mathbf{m^{\prime}}} ,
\end{equation}
where the operator
\begin{eqnarray}\label{L}
\mathcal{L}_{\mathbf{m},\mathbf{m^{\prime}}}=E\delta_{\mathbf{m},\mathbf{m^{\prime}}}
-\sum_{\mathbf{m^{\prime \prime}}}\delta_{\mathbf{m^{\prime
\prime}},\mathbf{m^{\prime}}} ,
\end{eqnarray}
is introduced for compactness (summation over
$\mathbf{m^{\prime\prime}}$ includes the nearest neighbours of the
site $\mathbf{m}$). The discrete-time equation (\ref{recursion
DN}) is a difference equation of the second order, which needs two
initial conditions (Cauchy problem for the difference equation).
The first natural condition is $\psi_{0,\mathbf{m}}=0$. The second
initial condition can be presented in the general form
$\psi_{1,\mathbf{m}}=\alpha_{\mathbf{m}}$ without any additional
conditions for the arbitrary constants $\alpha_{\mathbf{m}}$,
except that they are supposed to be finite. The rules for the
treatment of the field $\alpha_{\mathbf{m}}$ will be explained
below, section \ref{alpha}.

The recursion equation (\ref{recursion DN}) reveals a general
feature of the \textit{causality} - in its formal solution $\psi
_{n+1,\mathbf{m}}$ depends only on the random variables $\varepsilon
_{n^{\prime},\mathbf{m}^{\prime}}$ with $n^{\prime }\leq n$. This
gives us a hint for the \textit{exact solution} of the equation
provided that random variables $\varepsilon_{\mathcal{M}}$ on
different sites \textit{do not correlate}: $\left\langle
\varepsilon_{\mathcal{M}}
\varepsilon_{\mathcal{M}^{\prime}}\right\rangle \propto
\delta_{\mathcal{M},\mathcal{M}^{\prime}}$. In this case all
amplitudes $\psi$  on the r.h.s.\ of eq. (\ref{recursion DN}) are
statistically independent of $\varepsilon _{n,\mathbf{m}}$. In other
words, while performing the mathematical operations in eq.
(\ref{recursion DN}), e.g. \textit{taking the square} of both
equation sides, with a further averaging over an ensemble of
different realizations of the random potentials (symbol
$\left\langle \dots \right\rangle $), the average of the product of
amplitudes $\psi$ and potentials  $\varepsilon _{n,\mathbf{m}}$ can
be replaced by a product of the corresponding average quantities.
Therefore, the exact solution can be obtained only when the on-site
potentials $ \varepsilon_{n,m} $ are \textit{independently and
identically} distributed. We assume hereafter existence of the two
first moments, $\left\langle \varepsilon _{n,\mathbf{m}}
\right\rangle =0$ and $\left\langle \varepsilon
_{n,\mathbf{m}}^2\right\rangle =\sigma ^2$, where the parameter
$\sigma$ characterizes the \textit{disorder level}.

\subsection{Generalized diffusion}

The ensemble averaging using recursion eq.(\ref{recursion DN})
allows us to calculate averages of different functions containing
the amplitudes $\psi$. The problem arises, \textit{which} averages
we can and should calculate? E.g. several even moments of the
amplitude were calculated in the 1-D case \cite{Molinari}, whereas
only the second momentum was calculated for multidimensional
systems in \cite{Kuzovkov02,Kuzovkov04}. It is generally believed
that the \textit{complete information} is available from the
complete set of \textit{all} moments \cite{Pendry82}. However, it
is not clear, how this information should be analyzed. In
particular, how can the \textit{one} parameter of our interest -
the localization length $\xi$ - be extracted from the
\textit{infinite} set of amplitude momenta?

It was suggested in \cite{Comment} to calculate not only the
second, $\left\langle |\psi|^2 \right\rangle $, and other momenta,
but also \textit{physical values}, such as $\left\langle \ln|\psi|
\right\rangle $. The average quantities are divided in
\cite{Comment} into two categories: containing physical
information (e.g. $\left\langle \ln|\psi| \right\rangle $) and
containing no physical information (e.g. $\left\langle |\psi|^2
\right\rangle $).

In other words, one has to understand how the choice of a
semi-infinite system with a selected direction ($n$ axis) and the
boundary (layers in the transversal directions with $n=0,1$) allows
us to detect the localized states under question.

Let us consider for simplicity a 1-D system, where the amplitude
$\psi=\psi_n$ and the second initial condition $\psi_1=\alpha$.
Again, for \textit{simplicity} let us consider $\alpha$ to be a
real quantity, i.e. $|\psi|^2=\psi^2$. The recursive equation
reads:
\begin{eqnarray}\label{recursion1}
\psi_{n+1}=(E-\varepsilon_n)\psi_n-\psi_{n-1},
\end{eqnarray}
where $\left\langle \varepsilon_n \right\rangle = 0$,
$\left\langle \varepsilon^2_n \right\rangle = \sigma^2$.

Let us consider now the random walk problem (Brownian motion
\cite{Gardiner,Sokolov05} for the coordinate $\psi_n$) which is
described by the following equations
\begin{eqnarray}\label{recursion2}
\psi_{n+1}=\psi_n +\varepsilon_n .
\end{eqnarray}
There exists a mean squared displacement $\left\langle
\varepsilon^2_n \right\rangle = \sigma^2$ of the Brownian particle
during a step.

Considering $n$ as the discrete-time index, both eqs.
(\ref{recursion1}) and (\ref{recursion2}) describe the dynamics of
the system with a stochastic time-dependent perturbation
$\varepsilon_n$. The only difference is that the dynamics of the
system (\ref{recursion2}) is trivial: $\psi_n \equiv \psi_0$ when
there is no perturbation, $\varepsilon_n=0$. In contrast, the
dynamical system (\ref{recursion1}) even for $\varepsilon_n=0$
reveals proper dynamics (it is a second order equation!). In the
band region, $|E|< 2$, this corresponds to the bounded motion.
Therefore, the proper dynamics of both systems corresponds to the
\textit{bounded} trajectories, $\psi^2_n < \infty$.


Assuming now that $\sigma \neq 0$, eq. (\ref{recursion2}) describes
the \textit{normal diffusion}. The diffusion motion is characterized
by \textit{divergences}, e.g. for the mean time when the system
returns to the initial state. In particular, the momenta of the
amplitude $\psi_n$ are also divergent with the discrete-time $n$,
e.g.
\begin{equation}\label{normal diff}
\left\langle \psi^2_n \right\rangle = \psi^2_0+\sigma^2 n
\end{equation}
reveals a \textit{power law} divergence.
For
normal diffusion the mean square displacement $\left \langle
\psi^2_n \right\rangle $ of the Brownian particles is linear in
time.

To detect the diffusion, it is \textit{sufficient} to demonstrate
the divergence of the \textit{second moment} of the amplitude and to
establish its \textit{law of time-dependence}
\cite{Ebeling05,Kotomin96}. For example, for \textit{anomalous
diffusion} the mean quare displacement $\left \langle \psi^2_n
\right\rangle $ of the particles is no longer linear in time, but
follows a generalized Fick's second law \cite{Yuste04}.

Generally speaking, other moments contain additional information,
which, however, is not important. The divergence of the second
moment defines the conditions of the diffusion appearance:
\textit{if} the second moment is divergent, so are \textit{all other
even moments} (this is why the choice of a particular moment for
further analysis is \textit{not unique}). The \textit{law} of the
divergence of the time-dependence allows us to distinguish between
normal and abnormal diffusion. Important is the fact that the second
moment $\left\langle \psi^2_n \right\rangle $ in eq.
(\ref{recursion2}) can be found exactly analytically and thus we
prefer its use ($\psi^2$-definition). Speaking formally, the
diffusion could also be classified using other averaged quantities,
e.g. $\left\langle |\psi_n| \right\rangle $, but such a choice is
not convenient for mathematical reasons; it hinders an analytical
solution.

\begin{figure}[htbp]
  \begin{center}
    \epsfig{file=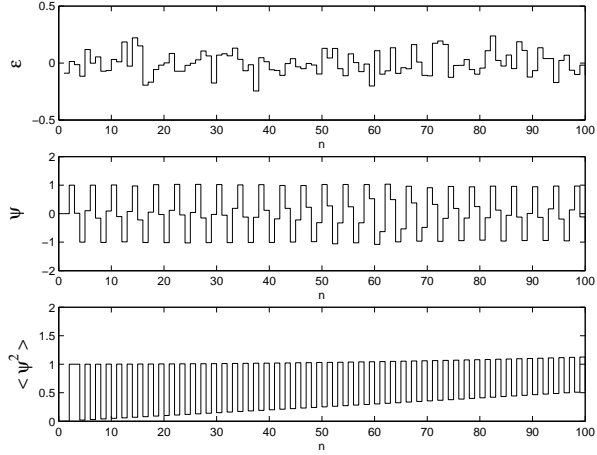,width=8cm}
    \caption{
     Exponential divergence of second moment for generalized diffusion,
     eq. (\ref{recursion1}), for $\sigma=0.1$ and $E=0$.
      }
    \label{fig: 2}
  \end{center}
\end{figure}

As is well-known, in a 1-D system described by eq.
(\ref{recursion1}) any disorder $\sigma \neq 0$ leads to
localization. This manifests itself in the simultaneous
divergence, as $n \rightarrow \infty$, of different average
quantities of $|\psi_n|$, e.g. the \textit{linear} divergence of
$\left\langle \ln |\psi_n| \right\rangle $ (log-definition of
localization), whereas the \textit{exponential} divergence occurs
for the powers of $|\psi_n|$ \cite{Molinari}. The appearance of
the localization in the approach based on eq. (\ref{recursion1})
is equivalent to the appearance of diffusion. In eq.
(\ref{recursion2}) the random perturbation $\varepsilon_n$ is
\textit{additive}, which determines the \textit{linear}
(power-law) character of the second moment divergence (normal
diffusion). In contrast, eq. (\ref{recursion1})  contains the
\textit{product} of $\varepsilon_n$ and $\psi_n$
(\textit{multiplicative noise}) which determines the
\textit{exponential} character of the divergence (see also
Brownian motion \cite{Qiang} and normal nonlinear diffusion
\cite{Munoz98} with multiplicative noise).

I.e., we can speak of the \textit{generalized diffusion},
analyzing the $\left\langle \psi^2_n \right\rangle = f(n)$
divergence as a function of $n$. The exponential localization here
corresponds to the exponential growth of $f(n) \propto
\exp(2\gamma n)$,  Fig. \ref{fig: 2}, where $\xi=1/\gamma$ can be
interpreted as the \textit{localization length} ($\psi^2$ -
localization definition \cite{Kuzovkov02,Kuzovkov04,Molinari}).
On the one hand, the numerical solution for the amplitude $\psi_n$
are plotted here using
eq.(\ref{recursion1})
together with arbitrary realiza\-tions of perturbations
$\varepsilon_n$ (two upper windows). On the other hand, the lower
window of Fig.\ref{fig: 2}
show the exact analytical solution for the second momenta
$\left\langle \psi^2_n \right\rangle $ (eqs.(\ref{s1D}) and
(\ref{z0}) of this paper).
The analysis of these results (as well as other data for different
$E$ and $\sigma$) demonstrates that purely exponential behaviour,
$\left\langle \psi^2_n \right\rangle \propto \exp(2\gamma n) $, is
valid only asymptotically, as $n \rightarrow \infty$. In other
words, the exponential behaviour of divergence is understood by
its definition, eq.(\ref{gamma2}). There is a transient regime for
finite $n$ values especially for small $\gamma$ (as shown in
Fig.\ref{fig: 2}) where the exponential behaviour is combined with
the oscillations corresponding to the proper dynamics,
eq.(\ref{recursion2}). Note that the bounded proper motion of
$\psi_n$ for $\varepsilon_n=0$ in the band region, $|E|< 2$,
corresponds to oscillating functions.

The $\psi^2$-definition is convenient not only because the
equations for the second moment can be exactly solved (see below).
It is important that this definition works also for the
non-exponential localization, which corresponds to the
non-exponential behaviour of $f(n)$. In contrast, the traditional
log-definition is valid only for the exponential localization, and
does not allow an exact analytical solution.

It was analytically shown \cite{Molinari} for 1-D systems that if
the higher moments of the random potential $\varepsilon_n$ can be
neglected (for $\sigma \rightarrow 0$) and we can restrict
ourselves to the only parameter $\sigma$, the exponents in
\textit{all} even moments (as well as in log-definition) are
proportional to each other, they differ only by numerical factors.
The proportionality of the
different exponents (Lyapunov exponents) in the exponential growth
indicates that different definitions of the generalized diffusion
are in fact \textit{identical} and one can use any of them, e.g
the $\psi^2$-definition.



Note that the emergence of the generalized diffusion phenomenon
and the existence of the phase diagrams in dynamical systems with
the proper dynamics  are analogous to \textit{critical phenomena}
in equilibrium and non-equilibrium systems. In this sense, it is
clear that use of the $\psi^2$-definition for the determination of
the phase-diagram is quite sufficient. Exactly solvable problems
in the physics of phase transitions \cite{Baxter}, e.g. the Ising
model, have demonstrated that in spite of the fact that the
physical quantities, i.e. long range order parameter, magnetic
susceptibility, specific heat, etc. are defined as
\textit{different averages}, they still behave \textit{similarly},
when only the phase-diagram is considered. This is not surprising
since one deals here with averages obtained for the \textit{same
statistical ensemble}. In particular, the critical temperature
obtained for the order parameter coincides with that obtained for
the susceptibility. This occurs for the \textit{exact} solutions,
although deviations can arise for the approximate solutions.
Similarly, the phase diagram for the system with a metal-insulator
transition can be defined uniquely if one determines any
non-trivial average quantity. This is why the choice of
$\left\langle \ln|\psi| \right\rangle $ is not convenient. A
product of a sum of variables can be represented as a polynomial,
where each term can be calculated, whereas the logarithm of a sum
cannot be represented this way and thus the causality principle
cannot be used. On the other hand, the averages of the powers of
the amplitude can be calculated exactly.

When we speak of the phase-diagram, we mean determination of the
boundaries of metallic and insulating phases. In this respect, the
results \cite{Kuzovkov02,Kuzovkov04} demonstrate that the second
order moments form a \textit{closed and linear} set of equations
and, where the only characteristic of the random potentials is the
second momentum, $\sigma^2$, have a simple interpretation. The
phase boundaries are defined by the only simple parameter $\sigma$
of the random potential. That is, calculation of second moments
(\textit{correlators}) is the \textit{direct way} to obtaining the
phase diagram.

\subsection{Correlators}

In general, the field $\alpha_{\mathbf{m}}$ is complex. We can
introduce the simplest non-trivial correlators based on the
$\psi^2$-definition:
\begin{eqnarray}
x(n)_{\mathbf{m},\mathbf{l}}=\left\langle \psi^{*}
_{n,\mathbf{m}}\psi
_{n,\mathbf{l}} \right\rangle , \\
y(n)_{\mathbf{m},\mathbf{l}}=\frac{1}{2}[ \left\langle \psi^{*}
_{n,\mathbf{m}}\psi _{n-1,\mathbf{l}} \right\rangle +\left\langle
\psi^{*} _{n-1,\mathbf{m}}\psi _{n,\mathbf{l}} \right\rangle ] .
\end{eqnarray}
Let us for \textit{mathematical simplicity} consider real
$\alpha_{\mathbf{m}}$. This is also justified by the fact that the
final eqs. (\ref{chi3}),(\ref{H}) remain the same. For real
$\alpha_{\mathbf{m}}$ definitions can be simplified:
\begin{eqnarray}
x(n)_{\mathbf{m},\mathbf{l}}=\left\langle \psi _{n,\mathbf{m}}\psi
_{n,\mathbf{l}} \right\rangle , \\
y(n)_{\mathbf{m},\mathbf{l}}=\left\langle \psi _{n,\mathbf{m}}\psi
_{n-1,\mathbf{l}} \right\rangle .
\end{eqnarray}
Taking square of the both sides of the eq. (\ref{recursion DN})
and using the \textit{causality principle}, one gets
\begin{eqnarray}\label{x(n)}
x(n+1)_{\mathbf{m},\mathbf{l}}=\delta_{\mathbf{m},\mathbf{l}}\sigma^2
x(n)_{\mathbf{m},\mathbf{m}}+x(n-1)_{\mathbf{m},\mathbf{l}}+\\
\nonumber
\mathcal{L}_{\mathbf{m},\mathbf{m^{\prime}}}x(n)_{\mathbf{m^{\prime}},\mathbf{l^{\prime}}}
\mathcal{L}_{\mathbf{l^{\prime}},\mathbf{l}}  -
\mathcal{L}_{\mathbf{m},\mathbf{m^{\prime}}}y(n)_{\mathbf{m^{\prime}},\mathbf{l}}-
\mathcal{L}_{\mathbf{l},\mathbf{l^{\prime}}}y(n)_{\mathbf{l^{\prime}},\mathbf{m}}
.
\end{eqnarray}
Analogously one obtains the additional relation
\begin{eqnarray}
y(n+1)_{\mathbf{m},\mathbf{l}}=-y(n)_{\mathbf{l},\mathbf{m}}+
\mathcal{L}_{\mathbf{m},\mathbf{m^{\prime}}}x(n)_{\mathbf{m^{\prime}},\mathbf{l}}
.
\end{eqnarray}
Note that eq.(\ref{x(n)}) contains explicitly the \textit{diagonal
correlators} with $\mathbf{l}=\mathbf{m}$:
$\chi(n)_{\mathbf{m}}=x(n)_{\mathbf{m},\mathbf{m}}$. It is
essential that diagonal correlators are \textit{positive} numbers,
$\chi(n)_{\mathbf{m}}\geq 0$.

\subsection{Z-transform}

It is convenient to use along the $n$ axis the Z-transform which
is common in discrete-time systems \cite{Weiss}:
\begin{eqnarray}
X(z)_{\mathbf{m},\mathbf{l}}=\sum_{n=0}^{\infty}\frac{x(n)_{\mathbf{m},\mathbf{l}}}{z^n}
, \\
Y(z)_{\mathbf{m},\mathbf{l}}=\sum_{n=0}^{\infty}\frac{y(n)_{\mathbf{m},\mathbf{l}}}{z^n}
.
\end{eqnarray}
After transformation one gets
\begin{eqnarray}
(z-z^{-1})X(z)_{\mathbf{m},\mathbf{l}}-\sigma^2\delta_{\mathbf{m},\mathbf{l}}\chi(z)_{\mathbf{m}}
-x(1)_{\mathbf{m},\mathbf{l}}=\\ \nonumber
\mathcal{L}_{\mathbf{m},\mathbf{m^{\prime}}}X(z)_{\mathbf{m^{\prime}},\mathbf{l^{\prime}}}
\mathcal{L}_{\mathbf{l^{\prime}},\mathbf{l}}  -
\mathcal{L}_{\mathbf{m},\mathbf{m^{\prime}}}Y(z)_{\mathbf{m^{\prime}},\mathbf{l}}-
\mathcal{L}_{\mathbf{l},\mathbf{l^{\prime}}}Y(z)_{\mathbf{l^{\prime}},\mathbf{m}} ,\\
zY(z)_{\mathbf{m},\mathbf{l}}=-Y(z)_{\mathbf{l},\mathbf{m}}+
\mathcal{L}_{\mathbf{m},\mathbf{m^{\prime}}}X(z)_{\mathbf{m^{\prime}},\mathbf{l}}
.
\end{eqnarray}
Hereafter the symmetry properties of the operator
$\mathcal{L}_{\mathbf{m},\mathbf{l}}$  are used.

The solution of the second equation is simple:
\begin{eqnarray}
Y(z)_{\mathbf{m},\mathbf{l}}=\frac{z}{z^2-1}
\mathcal{L}_{\mathbf{m},\mathbf{m^{\prime}}}X(z)_{\mathbf{m^{\prime}},\mathbf{l}}
-\frac{1}{z^2-1}X(z)_{\mathbf{m},\mathbf{l^{\prime}}}\mathcal{L}_{\mathbf{l^{\prime}},\mathbf{l}}
,
\end{eqnarray}
which leads to the equation for $X$-correlators
\begin{eqnarray} \label{X(z)}
(z-z^{-1})X(z)_{\mathbf{m},\mathbf{l}}-\sigma^2\delta_{\mathbf{m},\mathbf{l}}\chi(z)_{\mathbf{m}}
-x(1)_{\mathbf{m},\mathbf{l}}=
\frac{z^2+1}{z^2-1}\mathcal{L}_{\mathbf{m},\mathbf{m^{\prime}}}X(z)_{\mathbf{m^{\prime}},\mathbf{l^{\prime}}}
\mathcal{L}_{\mathbf{l^{\prime}},\mathbf{l}} \\
\nonumber-\frac{z}{z^2-1}[
\mathcal{L}_{\mathbf{m},\mathbf{m^{\prime}}}\mathcal{L}_{\mathbf{m^{\prime}},\mathbf{l^{\prime}}}
X(z)_{\mathbf{l^{\prime}},\mathbf{l}}+
X(z)_{\mathbf{m},\mathbf{m^{\prime}}}\mathcal{L}_{\mathbf{m^{\prime}},\mathbf{l^{\prime}}}
\mathcal{L}_{\mathbf{l^{\prime}},\mathbf{l}}].
\end{eqnarray}
Taking into account the second initial condition,
$x(1)_{\mathbf{m},\mathbf{l}}=\alpha_{\mathbf{m}}\alpha_{\mathbf{l}}$.

\subsection{Fourier transform}

The obtained equation can be easily solved formally via Fourier
expansion. Indeed, one gets for diagonal correlators
\begin{eqnarray}
\chi(z)_{\mathbf{m}}=\int
\frac{d^p\mathbf{k}}{(2\pi)^p}\chi(z,\mathbf{k})e^{-i\mathbf{k}\mathbf{m}}
,\\
\chi(z,\mathbf{k})=\sum_{\mathbf{m}}
\chi(z)_{\mathbf{m}}e^{i\mathbf{k}\mathbf{m}} .
\end{eqnarray}
A general equation for $X$-correlators has to be solved using
double Fourier expansion:
\begin{eqnarray}
X(z,\mathbf{k},\mathbf{k^{\prime}})=\sum_{\mathbf{m},\mathbf{l}}
X(z)_{\mathbf{m},\mathbf{l}}e^{i\mathbf{k}\mathbf{m}+i\mathbf{k^{\prime}}\mathbf{l}}
.
\end{eqnarray}
This leads to
\begin{eqnarray}\label{UX}
U(z,\mathbf{k},\mathbf{k^{\prime}})X(z,\mathbf{k},\mathbf{k^{\prime}})=
\alpha(\mathbf{k})\alpha(\mathbf{k^{\prime}})+\sigma^2
\chi(z,\mathbf{k}+\mathbf{k^{\prime}}) ,
\end{eqnarray}
where
\begin{eqnarray}
U(z,\mathbf{k},\mathbf{k^{\prime}})=(z-z^{-1})-
\frac{z^2+1}{z^2-1}\mathcal{L}(\mathbf{k})\mathcal{L}(\mathbf{k^{\prime}})+
\frac{z}{z^2-1}[\mathcal{L}^2(\mathbf{k})+
\mathcal{L}^2(\mathbf{k^{\prime}})] ,
\end{eqnarray}
and
\begin{equation}
\alpha(\mathbf{k})=\sum_{\mathbf{m}}
\alpha_{\mathbf{m}}e^{i\mathbf{k}\mathbf{m}} .
\end{equation}
The function
\begin{eqnarray}
\mathcal{L}(\mathbf{k})=E - 2\sum_{j=1}^p\cos (k_j)  \label{Lk}
\end{eqnarray}
arises as the Fourier transform of the operator
$\mathcal{L}_{\mathbf{m},\mathbf{l}}$.

\subsection{Diagonal correlators}

Assuming $\mathbf{k^{\prime
\prime}}=\mathbf{k}+\mathbf{k^{\prime}}$, eq.(\ref{UX}) can be
rewritten as
\begin{eqnarray}\label{UX2}
U(z,\mathbf{k},\mathbf{k^{\prime\prime}}-\mathbf{k})X(z,\mathbf{k},\mathbf{k^{\prime\prime}}-\mathbf{k})=
\alpha(\mathbf{k})\alpha(\mathbf{k^{\prime\prime}}-\mathbf{k})+\sigma^2
\chi(z,\mathbf{k^{\prime\prime}}) .
\end{eqnarray}
Using the relation
\begin{eqnarray}
\chi(z,\mathbf{k^{\prime \prime}})= \int
\frac{d^p\mathbf{k}}{(2\pi)^p}X(z,\mathbf{k},\mathbf{k^{\prime\prime}}-\mathbf{k})
,
\end{eqnarray}
eq.(\ref{UX2}) can be solved elementary for diagonal correlators:
\begin{eqnarray}\label{chi_k}
\chi(z,\mathbf{k})=\mathcal{H}(z,\mathbf{k})\chi^{(0)}(z,\mathbf{k}) ,\\
\chi^{(0)}(z,\mathbf{k})=\int
\frac{d^p\mathbf{k^{\prime}}}{(2\pi)^p}\frac{\alpha(\mathbf{k^{\prime}})\alpha({\mathbf{k}-\mathbf{k^{\prime}}})}
{U(z,\mathbf{k^{\prime}},\mathbf{k}-\mathbf{k^{\prime}})}
,\\
\frac{1}{\mathcal{H}(z,\mathbf{k})}=1-\sigma^2\int
\frac{d^p\mathbf{k^{\prime}}}{(2\pi)^p}\frac{1}
{U(z,\mathbf{k^{\prime}},\mathbf{k}-\mathbf{k^{\prime}})}
\label{Hk}.
\end{eqnarray}

Therefore we have obtained for an arbitrary second initial
condition (field $\alpha_{\mathbf{m}}$) an exact solution for
fundamental averages - \textit{diagonal correlators}. After use of
the Fourier transform the solution is expanded in modes labelled
by the $\mathbf{k}$-vector and describing the oscillations in the
solutions in the transversal direction. It is easy to see that the
$\psi^2$-definition leads to equations containing only the
$\sigma$ parameter of the random potentials. If $\sigma=0$, all
functions $\mathcal{H}(z,\mathbf{k})\equiv 1$ and thus
$\chi(z,\mathbf{k})\equiv \chi^{(0)}(z,\mathbf{k})$. In other
words, functions $\chi^{(0)}(z,\mathbf{k})$  correspond to
solutions in a \textit{completely ordered system}. Note that the
boundary conditions (field $\alpha_{\mathbf{m}}$) influence only
functions $\chi^{(0)}(z,\mathbf{k})$ which are independent of the
parameter $\sigma$. Introduction of disorder, $\sigma > 0$,
transforms the initial solution $\chi^{(0)}(z,\mathbf{k})$  into
$\chi(z,\mathbf{k})$, where the operator
$\mathcal{H}(z,\mathbf{k})$ describes this transformation.

In our derivation we assumed the system size in a transversal
direction to be infinite, $m_j\in(-\infty,\infty)$, and thus the
equations obtained correspond to the \textit{thermodynamic limit}
($L=\infty$). In the case of finite size of the system,
$m_j=0,1,\dots,L-1$ (cyclic conditions) the integrals should be
replaced by series. Let us consider for illustration the 2-D case:
$p=D-1\equiv 1$. Here the index $\mathbf{m}=m$ has $L$ values and
the number of modes $\mathbf{k}=k$ also equals $L$. That is,
analysis of the diagonal correlators leads to $L$-order matrices
(vectors). Surprisingly, this obvious fact was interpreted in ref.
\cite{Comment} that degrees of freedom of the initial problem were
lost in our paper \cite{Kuzovkov02}, thus questioning our proof of
the Anderson theorem. This conclusion was based on the traditional
approach of the transfer-matrix \cite{Comment,Pendry82} which
operates with much larger, $L^2 \times L^2$, matrices for 2-D. The
transition from $L^2 \times L^2$ matrices to $L$-order matrices was
interpreted as an error arising due to \textit{averaging over
initial conditions} \cite{Kuzovkov02} and thus artificially
introducing a translation in the transversal directions. However, we
do not use in the present study such an averaging (see also Section
\ref{Averaging}). The $L$-dimensional matrices are the
\textit{natural} mathematical tool for the description of diagonal
correlators. Moreover, an excessive size of $L^2 \times L^2$
matrices indicates that the transfer-matrix formalism is not
adequate. This is seen, in particular, from the existence of the
so-called \textit{trivial eigenvalues} of the transfer-matrix
\cite{Comment,Pendry82}, which are $\sigma$-independent. Moreover,
the transfer-matrix does not permit to make the transition to the
thermodynamic limit and thus to calculate the phase-diagram (for
more details see \cite{Reply}). These disadvantages are absent in
the 1-D case, when formally $L=0$ and the matrix dimensions
coincide. This is why we used in \cite{Kuzovkov02} the
transfer-matrix method only in the 1-D case and exclusively for the
illustration.

\section{Anderson localization and stability}

\subsection{Signals and filters}

The set of equations for the correlators is linear, the same is true
for eqs.(\ref{chi_k}) for each $\mathbf{k}$-mode. The modes are
\textit{normal} (no mixing of modes with different $\mathbf{k}$
values). The above mentioned divergence (exponential or
non-exponential growth) of the diagonal correlators, arising due to
localized states, transforms mathematically to the
\textit{instability} of the set of linear equations. For linear
discrete-time systems under study the most adequate formalism is
\textit{signal theory} \cite{Weiss}. Following
\cite{Kuzovkov02,Kuzovkov04,Weiss}, let us define
$\chi^{(0)}(z,\mathbf{k})$ as \textit{input signals}. In our
particular case this is a mathematical characteristic of the
$\mathbf{k}$-mode for an initial ordered systems. (Its sense is
explained below). The inverse Z-transform of a given mode can be
performed as follows:
\begin{equation}
\chi^{(0)}(z,\mathbf{k})\Rightarrow \chi^{(0)}(n,\mathbf{k}) .
\end{equation}
The input signal is a one dimensional numerical series. The
$\chi(z,\mathbf{k})$ is the \textit{output signal}; respectively
after the inverse Z-transform the latter corresponds to the
numerical series $\chi(n,\mathbf{k})$, associated with the
\textit{disordered system}, $\sigma > 0$. The relation between
these two signals is given by eq.(\ref{chi_k}): the output signals
are linear transformations of input signals performed through
$\mathcal{H}(z,\mathbf{k})$ functions called the \textit{system
filter} \cite{Weiss}.

Physical systems with time as independent variable are
\textit{causal} systems \cite{Weiss}. In our problem with
discrete-time index $n$ the causality is of primary importance for
the interpretation of the result. Note that in the general case
the inverse Z-transform is defined through the complex integral
\begin{equation}\label{inverse}
h(n,\mathbf{k})=\frac{1}{2\pi i}\oint \mathcal{H}(z,\mathbf{k})
z^n \frac{dz}{z} .
\end{equation}
The integration here is performed over the complex plane called
the \textit{region of convergence} (ROC) \cite{Kuzovkov02,Weiss},
well-defined for the causal filters. For the causal systems
$h(n,\mathbf{k})=0$ for $n<0$ holds, thus after the inverse
Z-transform eqs.(\ref{chi_k}) transform into the
\textit{convolution property} \cite{Weiss}:
\begin{equation}\label{inverse2}
\chi(n,\mathbf{k})=\sum_{l=0}^{n}h(n-l,\mathbf{k})\chi^{(0)}(l,\mathbf{k})
.
\end{equation}

One can see here the linear transformation of input signals into
output signals. The fundamental property of the signal theory is
that the divergence of the output signals is related to the
asymptotic behaviour of the filter coefficients $h(n,\mathbf{k})$
as $n \rightarrow \infty$, which is mathematically equivalent to
the \textit{poles} of the $\mathcal{H}(z,\mathbf{k})$ function of
the complex argument $z$ \cite{Weiss}.

\subsection{Fundamental mode}

Since the appearance of localized states leads to the divergence
of the diagonal correlator, it is necessary to clarify, which
\textit{fundamental} mode $\mathbf{k}=\mathbf{k_0}$  looses its
stability and thus is responsible for the divergence. Such a
problem is quite general in many branches of physics. However, in
our case it is quite obvious that the fundamental mode is
\textit{static}, $\mathbf{k_0=0}$.

The mode with $\mathbf{k\neq 0}$ describes transversal
oscillations of the diagonal correlators, but the \textit{sign} of
the divergent oscillating solution is \textit{not defined} (see
also section \ref{Averaging}). On the other hand, signs of the
diagonal correlators are well defined, they are
\textit{non-negative}, $\chi(n)_{\mathbf{m}}\geq 0$. That is, in
our particular case the solution is trivial, the fundamental mode
is \textit{static}. It is thus necessary to study the boundaries
of the stability of the fundamental mode $\mathbf{k_0=0}$ only
(the sign of the divergent non-oscillating solution is defined),
which means \textit{uniquely} the determination of the
\textit{phase diagram} of the system.

Assuming $\chi^{(0)}(z,\mathbf{0})\equiv S^{(0)}(z)$,
$\chi(z,\mathbf{0})\equiv S(z)$, $\mathcal{H}(z,\mathbf{0})\equiv
H(z)$), we arrive at
\begin{eqnarray}\label{chi2}
S(z)=H(z)S^{(0)}(z) .
\end{eqnarray}
After the inverse Z-transform one gets
\begin{equation}\label{inverse4}
s_n=\sum_{l=0}^{n}h_{n-l}s^{(0)}_{l} .
\end{equation}
The fundamental input signal $S^{(0)}(z)$ here and the fundamental
filter are defined by the integrals:
\begin{eqnarray}
S^{(0)}(z)= \frac{(z+1)}{(z-1)}\int\frac{d^p\mathbf{k}}{(2\pi)^p}
\frac{|\alpha(\mathbf{k})|^2}{[(z+1)^2/z-\mathcal{L}^2(\mathbf{k})]}
,\label{chi3} \\
\frac{1}{H(z)} = 1-\sigma^2
\frac{(z+1)}{(z-1)}\int\frac{d^p\mathbf{k}}{(2\pi)^p}
\frac{1}{[(z+1)^2/z-\mathcal{L}^2(\mathbf{k})]} .\label{H}
\end{eqnarray}
Note also the relation
\begin{equation}\label{inverse3}
h_n=\frac{1}{2\pi i}\oint H(z) z^n \frac{dz}{z} .
\end{equation}
That is, we have got here a new derivation of relations earlier
derived in Refs. \cite{Kuzovkov02,Kuzovkov04}. Note that for the
1-D case the integral disappears, since $p=0$, and one gets
\begin{eqnarray}
S^{(0)}(z)= \frac{(z+1)}{(z-1)} \frac{|\alpha|^2}{[(z+1)^2/z-E^2]}
,\label{chi3a} \\
\frac{1}{H(z)} = 1-\sigma^2 \frac{(z+1)}{(z-1)}
\frac{1}{[(z+1)^2/z-E^2)]} ,\label{Ha}
\end{eqnarray}

\subsection{Fundamental input signal}\label{alpha}

Let us discuss the \textit{physical sense} of the input signal
signal $S^{(0)}(z)$ (or $s^{(0)}_{n}$). Note that this corresponds
to an \textit{ideal} system with $\sigma=0$. When there is no
disorder, $\varepsilon_{\mathcal{M}}\equiv 0$, the particular
solutions of the tight-binding eq. (\ref{tight-binding}) are
\textit{bounded} functions, plane waves
$\exp(i\mathcal{K}\mathcal{M})$ with
$\mathcal{K}=\{k_1,k_2,\dots,k_D\}$, provided for a given energy
$E$
\begin{equation}\label{EK}
\mathcal{E}(\mathcal{K}) = E
\end{equation}
holds, where
\begin{equation}
\mathcal{E}(\mathcal{K}) = \sum_{j=1}^D 2 \cos(k_j) ,
\end{equation}
otherwise particular solutions are \textit{not bounded} functions,
which lie beyond the band and have no physical interpretation. If
we divide the wave vector into transversal or normal directions,
$\mathcal{K}\equiv \{\mathbf{k},k_D\}$, and keeping in mind that
\begin{equation}\label{EK2}
\mathcal{L}(\mathbf{k}) = 2\cos(k_D) ,
\end{equation}
for a given transversal mode $\mathbf{k}$ the bounded physical
solution exists, provided $|\mathcal{L}(\mathbf{k})| \leq 2$.
These relations are sufficient for understanding the physical
sense of the input signal.

Eq. (\ref{chi3}) contains in the integrand the function
\begin{eqnarray}\label{fun}
\frac{(z+1)}{(z-1)}\frac{1}{(z+1)^2/z-\mathcal{L}^2(\mathbf{k})}.
\end{eqnarray}
Under the condition $|\mathcal{L}(\mathbf{k})| \leq 2$ its inverse
Z-transform gives
\begin{eqnarray}
\frac{\sin^2(k_D n)}{\sin^2(k_D)} ,
\end{eqnarray}
where $k_D=k_D(\mathbf{k})$ is the solution of eq.(\ref{EK2}).
Assuming that the arbitrary field $\alpha_{\mathbf{m}}$ is
selected in such a way that the Fourier transform coefficients
$\alpha(\mathbf{k})$ are not zero \textit{only} if
$|\mathcal{L}(\mathbf{k})| \leq 2$, eq.(\ref{chi3}) after the
Z-transform reads
\begin{eqnarray}
s^{(0)}_{n}= \int\frac{d^p\mathbf{k}}{(2\pi)^p}
\frac{|\alpha(\mathbf{k})|^2 \sin^2(k_D n)}{\sin^2(k_D)} .
\label{chi4}
\end{eqnarray}
The $s^{(0)}_{n}$ is \textit{bounded} for any $n$.

On the other hand, assuming that $|\mathcal{L}(\mathbf{k})| \leq
2$ is violated for certain $\mathbf{k}$, so the
$\alpha(\mathbf{k})\neq 0$ for $|\mathcal{L}(\mathbf{k})| \geq 2$,
the inverse Z-transform of eq.(\ref{fun}) leads to the function
\begin{eqnarray}
\frac{\sinh^2(\kappa_D n)}{\sinh^2(\kappa_D)} ,
\end{eqnarray}
which increases without bounds as $n \rightarrow \infty$. Here
$\kappa_D=\kappa_D(\mathbf{k})$ is the solution of equation
\begin{equation}\label{Ekappa}
|\mathcal{L}(\mathbf{k})| = 2\cosh(\kappa_D) .
\end{equation}
The input signal $s^{(0)}_{n}$ is also increasing to infinity as a
function of $n$. Note that both input and output signals are
\textit{real} values.

In other words, on can conclude that any \textit{physical
solution} of an ideal system ($\sigma=0$) with \textit{bounded}
wave functions has a one-to-one correspondence to the mathematical
object - a \textit{bounded} 1-D input signal $s^{(0)}_{n}$. Such
solutions always exist, if the energy $E$ lies within the band
interval. On the other hand, formal solutions without physical
interpretation (\textit{unbounded} solutions beyond the band)
correspond to the 1-D \textit{unbounded} input signals. Note that
particular numerical values of input or output signals are not
important from the point of view of the signal theory \cite{Weiss}
whose main concern is a \textit{qualitative discrimination}
between bounded and unbounded signals.

\subsection{Fundamental filter}

The quantity characterizing the phase diagram of a disordered
system is the \textit{fundamental filter}, eq.(\ref{H}). Its idea
is quite clear \cite{Kuzovkov02,Kuzovkov04}. Let us consider the
physical states inside the band, $|E|<2D$. All these states are
described by wave functions bounded in amplitude, which
corresponds to a bounded 1-D input signal $s^{(0)}_{n}$. In a
disordered system this signal transforms into $s_{n}$, according
to eq. (\ref{inverse4}). This output signal $s_{n}$ can be either
bounded, or infinitely growing (generalized diffusion).

The natural interpretation of \textit{bounded output signals} is
that they - as before - correspond to the \textit{delocalized
states} bounded in amplitude. The \textit{unbounded output
signals} correspond to the \textit{localized states},
respectively, which lead in the semi-infinite system to the
divergence of the diagonal correlators. An important result of
signal theory \cite{Weiss} is that the divergence does not result
from properties of a particular bounded input signal; the cause
lies in the filter, $H(z)$ or $h_n$. We define the phase diagram
of the Anderson model based on a general concept of the signal
theory known as a \textit{BIBO stability} \cite{Weiss}. Namely, a
system is BIBO stable if every \textbf{B}ounded \textbf{I}nput
leads to a \textbf{B}ounded \textbf{O}utput. A stable system
(delocalized states) is characterized by a \textit{stable} filter
$H(z)$; its main property is the \textit{absence of poles} in the
complex $z$-plane outside the circle  $|z|>1$. An
\textit{unstable} system (localized states) is characterized by
the \textit{unstable} filter $H(z)$ \textit{with poles} outside
the circle $|z|>1$. As was shown in \cite{Kuzovkov02,Kuzovkov04},
in the Anderson model the pole $z_0$ of the unstable filter lies
on the real axis. Consequently, $z_0=\exp(2\gamma)$, where
$\gamma$ is the Lyapunov exponent ($\psi^2$-definition)
\cite{Kuzovkov02,Kuzovkov04}. Using the inverse Z-transform, one
can easily find that asymptotically $h_n \sim \exp(2\gamma n)$,
i.e. it is divergent. The localization length is defined as
$\xi=\gamma^{-1}$.

The mathematical formalism of the signal theory allows to extend
and comple\-ment this result for the energy range $|E|>2D$, where
in the absence of disorder there existed only mathematical
(divergent) solutions having no phy\-sical interpre\-tation. It is
known that disorder extends the band, i.e. new physical states
arise also at $|E|>2D$. The question is: can delocalized states
exist amongst these new states? The answer is simple. In the
region $|E|>2D$ only \textit{unbounded input signals}
$s^{(0)}_{n}$ exist. Such a signal cannot be transformed into a
\textit{bounded output signal} $s_{n}$. The output signal $s_{n}$
is always unbounded with a dual interpretation: on the one hand it
corresponds to mathematical solutions (no physical
interpretation); on the other hand, the divergence of the output
signal can be associated with an emergence of the localized
states. I.e., an emergence of delocalized states in the region
$|E|>2D$ (outside the band) is impossible \cite{Kuzovkov04}.

Use of the filter function $H(z)$ is a general and abstract method
for describing the metal-insulator transition, valid for any space
dimension. Earlier \cite{Kuzovkov02} we used a less fundamental
approach for the 1-D problem, based on the calculation of the
Lyapunov exponent for the $\psi^2$-definition
\begin{equation}\label{gamma2}
\gamma=\lim_{n \rightarrow \infty} \frac{1}{2n} \ln s_n ,
\end{equation}
where for the 1-D case $s_n \equiv \left\langle \psi_n^2
\right\rangle$. Taking into account that the divergence of the
output signal $s_n$ is caused by the divergence of the filter
$h_n$, one can use also the following equation
\begin{equation}\label{gamma3}
\gamma=\lim_{n \rightarrow \infty} \frac{1}{2n} \ln h_n .
\end{equation}
In the 1-D case both definitions of $\gamma$ are equivalent, since
for a fixed energy and disorder, for a second initial condition
$\psi_1=\alpha$,  one gets the \textit{single} output signal
$s_n$.

Let us use for illustration eqs.(\ref{chi3a}), (\ref{Ha}).
Restricting ourselves by the band center $E=0$ (which simplifies
the equations) and making the inverse Z-transform, one gets
\begin{eqnarray}
s_n=|\alpha|^2 \frac{z_0}{(z^2_{0}+1)}[z^n_0-(-1/z_0)^n] \label{s1D},\\
h_n=\delta_{n,0}+ \frac{(z^2_0-1)}{(z^2_{0}+1)}[z^n_0-(-1/z_0)^n]
, \label{h1D}\\
z_0=\frac{\sigma^2+\sqrt{4+\sigma^4}}{2} \label{z0},
\end{eqnarray}
provided $s^{(0)}_n=|\alpha|^2 [1-(-1)^n]/2$ (bounded input
signal). Indeed, simultaneous divergence of the output signal
$s_n$ and the filter $h_n$ arises due to the fact that the
asymptotic behaviour of both quantities is determined by the same
parameter, $z_0=\exp(2\gamma)>1$.

However, this equivalence is no longer valid for a space dimension
higher than one. Since the second initial condition is defined by
the field $\alpha_{\mathbf{m}}$, this corresponds to a continuum
of input signals $s^{(0)}_n$ and, respectively, a continuum of
output signals $s_n$. That is, the definition (\ref{gamma2}) is
not valid, since it is not clear that different signals should
correspond to the same Lyapunov exponent $\gamma$. Formally, a
whole continuum of solutions should be analyzed. However,
eq.(\ref{inverse4}) demonstrates that the fundamental Lyapunov
exponent $\gamma$ does \textit{not depend} at all on the field
$\alpha_{\mathbf{m}}$, it is sufficient to define the Lyapunov
exponent using only eq.(\ref{gamma3}).

\subsection{Phase diagram and multiplicity of solutions}

A more detailed analysis \cite{Kuzovkov02,Kuzovkov04,Reply}
reveals another problem of the definition (\ref{gamma2}): this is
valid \textit{only} in the 1-D case where for any disorder
$\sigma$ only the phase of the localized solutions exists.

As it was mentioned above, for the calculation of the fundamental
filter $h_n$ using the inverse Z-transform, eq.(\ref{inverse3}),
the contour integration over the so-called \textit{region of
convergence} (ROC) \cite{Weiss} is necessary, provided the filter
under consideration is causal. It is shown
\cite{Kuzovkov02,Kuzovkov04} that dependent on the energy $E$ and
disorder $\sigma$, two general cases are possible for the Anderson
model with $D \geq 2$. In the first case, the ROC lies
\textit{outside} the circle $|z|>1$. The filter $h_n$ is thus
\textit{unstable} and describes the localized states (insulating
phase). In the second case, the ROC consists of \textit{two
domains} in the complex plane: one domain is \textit{inside} the
circle $|z|\leq 1$, another one - \textit{outside} this circle,
$|z|>1$. Respectively, there are \textit{two ways} to calculate
the integral using eq. (\ref{inverse3}). The double solution
arises, one describing delocalized states (filter $h^{(-)}_n$,
metallic phase), the other localized states (filter $h^{(+)}_n$,
insulating phase).

Consequently, in this range of parameters $E$ and $\sigma$ the
\textit{two phases} can co-exist, this is why the metal-insulator
transition in the Anderson model has to be analyzed in terms of
\textit{first-order} phase transition theory
\cite{Kuzovkov02,Kuzovkov04,Reply}. It is known that complex
(first or higher order) phase transitions introduced purely by
noise can be generated for spatially extended systems
\cite{VandenBroeck}. For arbitrary random potentials the wave
function can be \textit{either} localized \textit{or} delocalized
(no co-existence!). However, in terms of \textit{statistics} of an
ensemble of random potentials, \textit{both} localized
\textit{and} delocalized solutions can arise with a
\textit{comparable probability}. Namely this comparability of
probabilities is the main characteristic of the first-order phase
transition, in contrast to the second-order transitions where
either a pure metallic phase (no localized states) or a pure
insulating phase (no delocalized states) should exist:
co-existence is impossible.



\begin{figure}[htbp]
  \begin{center}
    \epsfig{file=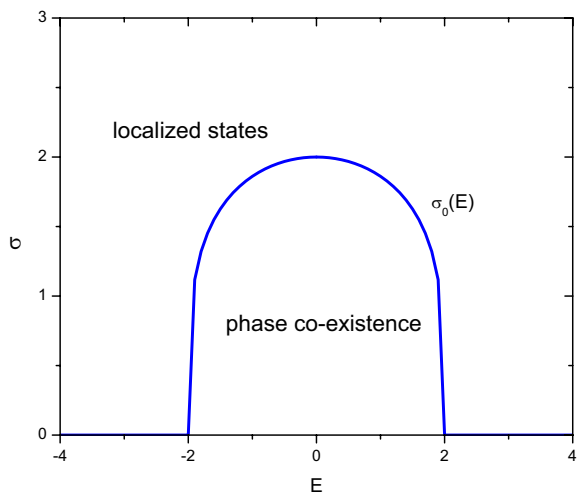,width=10cm}
    \caption{
     Phase diagram for 2-D case.
      }
    \label{fig: 3}
  \end{center}
\end{figure}
For the illustration let us discuss here the phase diagram in
terms of energy $E$ and disorder $\sigma$ for 2-D case as shown in
Fig.\ref{fig: 3} (the band region $|E| \leq 4$). The other
dimensions are discussed in Ref.\cite{Kuzovkov04}. The diagram in
Fig.\ref{fig: 3} shows that  all states with energies $|E| > 2$
are localized at arbitrarily weak disorder. The localization at
the finite disorder parameter $\sigma \geq \sigma_{0}(E)$, where
according to Ref.\cite{Kuzovkov02}
\begin{equation}\label{Esigma}
\sigma_{0}(E) = 2 (1-E^2/4)^{1/4} ,
\end{equation}
is also observed in the energy range $\left| E\right| \leq 2$. In
other words, only pure insulating phase (no delocalized states)
takes place in these cases. The localized states are observed also
in the rest of the phase diagram, namely, for small energies,
$\left| E\right| \leq 2$, and small disorder, $\sigma <
\sigma_{0}(E)$. However, as it was mentioned above, the localized
states could be observed only for a certain realization of the
random potentials. For other realizations the delocalized states
occur in the same range of energy and disorder, $\left| E\right|
\leq 2$ and $\sigma < \sigma_{0}(E)$.

Phase co-existence creates a serious problem of the choice of an
adequate mathematical formalism. Strictly speaking, when
calculating the average quanti\-ties on the ensemble of different
realizations of random potentials, the contri\-butions from pure
metallic and insulating phases are considered as equivalent.
However, such \textit{heterophase averages} have no physical sense
\cite{Kuzovkov02,Kuzovkov04,Reply}. In a two-phase system with
first-order phase transition one is interested in properties of
\textit{pure phases}. From this point of view, the earlier
introduced output signals $s_n$  are also heterophase averages.
This is why the derivation of equations for signals is not a goal,
but the tool for obtaining a more fundamental property - the
filter $H(z)$ (or $h_n$). As was noted
\cite{Kuzovkov02,Kuzovkov04,Reply}, it is such a filter which
reveals the \textit{multiplicity of solutions} and which permits
us to determine the phase diagram for the Anderson model.

\subsection{Averaging over initial conditions} \label{Averaging}

Let us discuss now the problem of \textit{averaging over initial
conditions} \cite{Kuzovkov02,Kuzovkov04}. This procedure was
introduced in Ref. \cite{Kuzovkov02}, its meaning is quite simple.
First of all, we choose a particular second initial condition, i.e.
a field $\alpha_{\mathbf{m}}$. Secondly, let us consider the field
$\alpha^{\prime}_{\mathbf{m}}\equiv \alpha_{\mathbf{m+m_0}}$
obtained from the first field as the result of a trivial
\textit{translation} in transversal direction by the vector
$\mathbf{m_0}$. It is obvious that the relevant diagonal correlators
$\chi^{\prime}(n)_{\mathbf{m}} \equiv \chi(n)_{\mathbf{m+m_0}}$ can
also be obtained by the \textit{argument shift} in a transversal
direction. Both solutions are \textit{physically equivalent}. The
Fourier transform of the field $\alpha_{\mathbf{m}}$ with the vector
shift $\mathbf{m_0}$ satisfies a simple relation:
$\alpha^{\prime}(\mathbf{k})\equiv
\exp(-i\mathbf{k}\mathbf{m_0})\alpha(\mathbf{k})$.

Eqs. (\ref{chi_k}) to (\ref{Hk}) clearly demonstrate that the
vector shift affects only signals for particular modes
$\mathbf{k}$, and all physically equivalent solutions differ from
each other only by \textit{phases}:
\begin{eqnarray}
\chi^{\prime(0)}(z,\mathbf{k}) \equiv
\exp(-i\mathbf{k}\mathbf{m_0})\chi^{(0)}(z,\mathbf{k}) , \\
\chi^{\prime}(z,\mathbf{k}) \equiv
\exp(-i\mathbf{k}\mathbf{m_0})\chi(z,\mathbf{k}) .
\end{eqnarray}
Note that only the fundamental mode $\mathbf{k=0}$ remains
invariant. Averaging now the signal over all translations in
transversal directions (averaging over initial conditions
\cite{Kuzovkov02}) gives zero for all non-fundamental modes,
$\mathbf{k\neq 0}$: $\overline{\chi^{(0)}(z,\mathbf{k}) }=0$,
$\overline{\chi(z,\mathbf{k}) }=0$. Nonzero is only the
fundamental mode ($\mathbf{k\equiv 0}$):
$\overline{\chi^{(0)}(z,\mathbf{0}) }=S^{(0)}(z)$,
$\overline{\chi(z,\mathbf{0}) }=S(z)$. The average diagonal
correlator loses its dependence on the argument $\mathbf{m}$:
$\overline{\chi(n)_{\mathbf{m}}}=s_n$.  That is, averaging over
initial conditions is an efficient tool for getting rid of all
non-fundamental modes which are non-essential for the analysis.

Note that eq. (\ref{chi3}) contains
$\Gamma(\mathbf{k})=|\alpha(\mathbf{k})|^2$ which is a function of
the field $\alpha_{\mathbf{m}}$, its Fourier transform is
$\Gamma_{\mathbf{m}}$. The averaging over initial conditions
corresponds mathematically in the initial eq. (\ref{X(z)}) to the
replacement of the initial condition
$x(1)_{\mathbf{m},\mathbf{l}}=\alpha_{\mathbf{m}}\alpha_{\mathbf{l}}$
by the average,
$\overline{x(1)_{\mathbf{m},\mathbf{l}}}=\Gamma_{\mathbf{m-l}}$.
After such a \textit{replacement} of initial conditions the system
becomes much simpler: it is \textit{translation-invariant} in
transversal directions which makes use of the double Fourier
transform unnecessary. In other words, averaging over initial
conditions is nothing but a simple mathematical trick which
permits to get quickly the fundamental property of system - the
filter $H(z)$. Such tricks are based on the fact that the filters
are universal system characteristics, describing the solution
transformation from order to disorder. Such universal
characteristics are \textit{independent} of the initial conditions
and other details. The initial conditions determine such
non-universal properties as signals. Consequently, in order to
obtain the fundamental filter $H(z)$ and then the phase diagram,
one can perform \textit{linear operations} with signals, in
particular, averaging over initial conditions.

Another example of such operations is the following. Let us
consider the field $\alpha_{\mathbf{m}}$ not as fixed (second
initial condition), but as a random variable, with simultaneous
averaging \textit{over the field} $\alpha_{\mathbf{m}}$ and
\textit{over the ensemble of the random potential realizations}
$\varepsilon_{\mathcal{M}}$. We assume the absence of correlations
between the potentials $\varepsilon_{\mathcal{M}}$ and the field
$\alpha_{\mathbf{m}}$; unlike the correlation between the field
components characterized by the arbitrary correlation function
$\left\langle \alpha_{\mathbf{m}} \alpha_{\mathbf{l}}\right\rangle
=\Gamma_{\mathbf{m-l}}$. The initial condition in eq.(\ref{X(z)})
becomes $x(1)_{\mathbf{m},\mathbf{l}}=\Gamma_{\mathbf{m-l}}$.
Taking into account that the operator
$\mathcal{L}_{\mathbf{m},\mathbf{m^{\prime}}}$ by its definition,
eq.(\ref{L}), depends only on the on the argument difference,
$\mathbf{m}-\mathbf{m^{\prime}}$, one gets the translation
invariant system, where $X(z)_{\mathbf{m},\mathbf{l}} \equiv
\hat{X}(z)_{\mathbf{m-l}}$, provided for the diagonal correlators
$X(z)_{\mathbf{m},\mathbf{m}}=\chi(z)_{\mathbf{m}} \equiv
\hat{X}(z)_{\mathbf{0}} = S(z)$ holds (the diagonal correlators
are $\mathbf{m}$-independent).

In this case in order to solve eq. (\ref{X(z)}), it is sufficient
to use a single Fourier transform
\begin{eqnarray}
\hat{X}(z)_{\mathbf{m}}=\int
\frac{d^p\mathbf{k}}{(2\pi)^p}\hat{X}(z,\mathbf{k})e^{-i\mathbf{k}\mathbf{m}}
.
\end{eqnarray}
Instead of eq.(\ref{UX}) one gets
\begin{eqnarray}\label{UX3}
U(z,\mathbf{k},\mathbf{-k})\hat{X}(z,\mathbf{k})=
\Gamma(\mathbf{k})+\sigma^2 S(z) ,
\end{eqnarray}
which gives
\begin{eqnarray}
S(z)=\int \frac{d^p\mathbf{k}}{(2\pi)^p}\hat{X}(z,\mathbf{k}) .
\end{eqnarray}
As a result, one returns to the fundamental eqs.(\ref{chi2}) to
(\ref{H}), with replacement of the $|\alpha(\mathbf{k})|^2$ for
$\Gamma(\mathbf{k})$.

\section{Conclusion}

We would like to stress that in this paper we presented a
mathematically rigorous method for the calculation of the phase
diagram for the Anderson localization in arbitrary dimensions,
which was briefly discussed earlier \cite{Kuzovkov02,Kuzovkov04}.
The phase diagram for the metal-insulator transition is obtained
using the Lyapunov exponent $\gamma$. Localized states correspond
to values of $\gamma > 0$,  i.e. a divergence of the averages over
wavefunctions. This divergence is mathematically similar to the
divergence of averages for the diffusion motion. That is,
transition to the localized states can be treated as a generalized
diffusion. From this viewpoint, in order to determine the range of
the existence of localized states (i.e. the phase diagram) and the
type of localization (exponential or non-exponential), it is
sufficient to solve equations for the joint correlators. We have
shown that these equations are \textit{exactly} solvable
analytically.

In its turn, the appearance of the generalized diffusion arises
due to the instability of a fundamental mode corresponding to
correlators. The generalized diffusion can be described in terms
of signal theory, which operates with the concepts of input and
output signals and the filter function. Delocalized states
correspond to bounded output signals, and localized states to
unbounded output signals, respectively. Transition from bounded to
unbounded signals is defined uniquely by the filter function
$H(z)$, or more precisely, by  the position of its poles in the
complex plane. This function can be calculated for arbitrary space
dimension D.

\ack{V.N.K. gratefully acknowledges the support of the Deutsche
Forschungsgemeinschaft. The authors are indebted to E. Kotomin for
detailed discussions of the paper.}


\end{document}